# Complex wavefront engineering via decoupled space-time modulation

Virat Tara[1], Anna Wirth-Singh[2], Johannes E. Fröch[1] and Arka Majumdar[1,2]

[1]Department of Electrical and Computer Engineering, University of Washington, Seattle, WA, USA

[2]Department of Physics, University of Washington, Seattle, WA, USA

Email: vtara@uw.edu, arka@uw.edu

## Abstract

Solid-state Spatial Light Modulators (SLMs) are fundamentally limited in their ability to achieve high spatial complexity and high temporal bandwidth simultaneously. High-speed, low-energy modulation requires sub-wavelength active mode volumes, and sophisticated spatial wavefront engineering necessitates an ultra-fine pixel pitch. While small pixels can simultaneously solve both, in conventional architectures, the dense 2D electrical routing required for such pixels creates an insurmountable physical bottleneck. This results in a compromise between the SLM refresh rate, number of pixels and the field of view. Here, we demonstrate a hybrid architecture that overcomes this limit by spatially decoupling the electrical modulation plane from the optical output plane. By integrating a metasurface doublet with a photonic integrated circuit (PIC)-based optical phased array (OPA), we achieve independent 2D electrical control over each phase-element while simultaneously realizing a three-fold reduction in effective pixel pitch. This decoupling allows us to maintain the small active volume required for high-speed operation, while circumventing the routing constraints of dense spatial array of emitters. We utilize this platform to demonstrate tunable varifocal lensing, 2D beam steering, and 2D holography. Our work provides a scalable foundation for next-generation solid-state SLMs that simultaneously offer high speed, low power consumption, and large field of view.

## Introduction

Electrically tunable metasurfaces and photonic integrated circuits (PIC) offer a highly promising pathway toward fully reconfigurable planar optics[1], [2], [3] serving as a potential foundation for next-generation solid-state spatial light modulators (SLMs). By eliminating moving parts, these devices overcome the limitations of their mechanical counterparts, offering significantly higher speeds and greater robustness. Solid-state SLMs hold the promise to advance applications spanning light detection and ranging[4], sensing[5], optical computing[6], [7], quantum computing[8], [9], space communication and even biomedicine for deep tissue imaging[10] and brain-machine interfaces[11]. While significant progress has been made in electrically reconfigurable metasurfaces using platforms such as phase-change materials[12], [13], [14], [15], [16], [17], [18], [19], electro-optic materials[20], [21], [22], [23], liquid crystals[24], [25], optofluidics[26], and degenerately doped semiconductors, e.g., ITO[27], contemporary devices remain fundamentally constrained by only 1D electrical control, limited field of view (FOV), high power consumption, and restricted modulation speeds due to lack of individual control over each pixel.

The requirements for an ideal SLM are twofold: it must possess a wide, aliasing-free field of view (FOV) using efficient space modulation, and it must enable independent, high-speed time modulation of every pixel. The first requirement is satisfied by minimizing the pixel pitch, a challenge that has been largely resolved by the advent of metasurfaces[28], [29], which typically arrange meta-atoms in sub-wavelength lattices to suppress diffraction artifacts (detailed in Supplementary Information). Similarly, ultra-high speed modulation up to GHz of isolated individual pixels has been widely demonstrated[16], [18]. However, a fundamental integration bottleneck emerges when attempting to satisfy both requirements simultaneously. Although scaling down the pixel pitch reduces the per-pixel power consumption, it makes it physically prohibitive to maintain independent 2D electrical control over every pixel. In fact, the global system refresh rate ($f_R$) is often significantly slower than the single pixel modulation rate ($f_{pixel}$). To understand the severity of this limitation, it is instructive to examine how current SLMs attempt to circumvent this packaging challenge.

To circumvent the electrical addressing challenge, commercially available SLMs frequently employ active-matrix refresh mechanisms. In this architecture, an integrated memory (such as static random-access memory) holds the pixel's state between refresh cycles, similar to displays in consumer electronics. Yet, this approach introduces its own limitations: the bulk of the added memory circuitry precludes achieving a true sub-wavelength pitch, and the overall modulation speed is severely bottlenecked by the charging and discharging times of the integrated capacitor.

To achieve independent 2D control over pixel modulation, many state-of-the-art SLMs rely on optical addressing[30], [31], [32]. Although optical addressing is intuitively associated with ultrafast and highly parallel operation, the foundational control signal must still originate electrically. As a result, utilizing optical control essentially reduces to driving the primary SLM with a secondary SLM. For instance, the system in [26] utilizes a commercial digital micromirror device (DMD) to optically address the primary metasurface SLM; thus, the mechanical limitations of the DMD ultimately bottleneck the speed and performance of the entire system. Similarly, the authors in [27] employ a static mask to modulate the illumination source prior to their SLM interaction. In another approach, the work in [25] utilizes a 100 μm pitch micro-LED array controlled by a complementary metal-oxide-semiconductor (CMOS) drive chip to modulate a photonic crystal SLM via free-carrier injection. While free-carrier dynamics can operate in the > 100 MHz regime, the micro-LED array is unable to deliver sufficient energy to induce the required phase shift to make holograms when operating at those frequencies. Furthermore, although an isolated micro-LED pixel can achieve speeds of 145 MHz, driving even four LEDs simultaneously introduces severe electrical crosstalk. This interference imposes a 2.88 dB power penalty, meaning nearly twice as much optical power is required at the receiver to overcome the signal degradation[33]. Thus, even though the primary SLM is optically addressed, its operation is limited by the electrical crosstalk when operating more than one LED at a given time.

Therefore, for both optical and direct electrical addressing schemes, the challenge fundamentally hinges on routing electrical control signals to individual pixels without increasing the pitch or pixel size. Given the difficulty of this challenge, most of the electrically reconfigurable SLM works have shown only 1D or single pixel electrical control[13], [14], [15], [16], [17], [18], [20], [21], [22], [23], [24], [25], [26], [27]. A crossbar control architecture can overcome these challenges however it suffers from severe inter-pixel crosstalk due to current leakage. Therefore, it is mostly suitable to be used with slow and non-volatile phase change materials[34], making it suitable for only display applications.

Similar to their free-space metasurface counterparts – PIC-based SLMs or Optical Phased Arrays (OPA) generally struggle to achieve true reconfigurable 2D holography due to their large foot print as compared to metasurfaces, which requires complicated fabrication and electrical addressing schemes to overcome[35], [36]. Furthermore, existing approaches lack a clear pathway for scaling OPA-based SLMs, fundamentally restricting most experimental demonstrations to devices with a limited number of emitters[37]. To circumvent these challenges, many PIC based systems default to 1D or 2D beam scanning[3], [38], [39], [40], [41], [42], [43] where 2D scanning is achieved by electrical control in one direction and laser scanning in the other direction. Furthermore, beam scanning also requires averaging at the detector thereby reducing the refresh rate ($f_R$) of the SLM which makes it also suitable only for display applications. While a recent work proposed an innovative approach (by exploiting spectral filtering) to build high-speed SLMs[44], the optical path length running into meters required by the SLM creates a prohibitive physical footprint, severely limiting its practical applications.

To overcome the aforementioned challenges, we propose a hybrid architecture that decouples the space and time degrees of freedom. By spatially decoupling the electrical control layer (hereafter referred to as the temporal plane) from the optical output plane (hereafter referred to as the spatial plane), we relax the stringent spatial constraints associated with sub-wavelength pixel pitches, providing ample area to route electrical control signals to individual pixels. Specifically, we employed a PIC as the temporal plane, owing to their tight light confinement and subsequent low modulation energy (see Supplementary Information on the power consumption of a SLM through the temporal-space bandwidth product). The outcoupled light of the PIC has a large pitch, dictated by the separation between gratings. To compress the beam pitch in the spatial plane, we used a metasurface doublet (MS1 and MS2) directly atop the PIC, with MS2 serving as the final optical output plane. Using this hybrid PIC and metasurface platform, we demonstrate a 2D electrically addressable phase SLM capable of dynamic varifocal lensing, beam steering, and holography. Giving us the ability to perform 3D light manipulation while keeping the overall system's refresh rate same as the individual pixel's maximum modulation rate ($f_R = f_{pixel}$). To the best of our knowledge, this work represents the first successful demonstration of these combined functionalities in a PIC-based SLM. Moreover, due to the space-time decoupling, we can use simple PIC components without the need to design complex structures to minimize the size of various PIC components such as multi-mode interference (MMIs) and grating couplers. Our work lays the foundation for next-generation solid-state SLMs offering high speed, low power consumption, and large FOV with sub-wavelength pixel pitch. We also present a comparison table for state-of-the-art reconfigurable SLMs in Supplementary Information. We note that a recent study also proposes combining photonic integrated circuits (PICs) and metasurfaces to enable wide-field-of-view SLMs. However, because their architecture relies exclusively on amplitude modulation, the platform is fundamentally restricted to low-resolution, highly discrete beam steering[45].

**Results and Discussions**

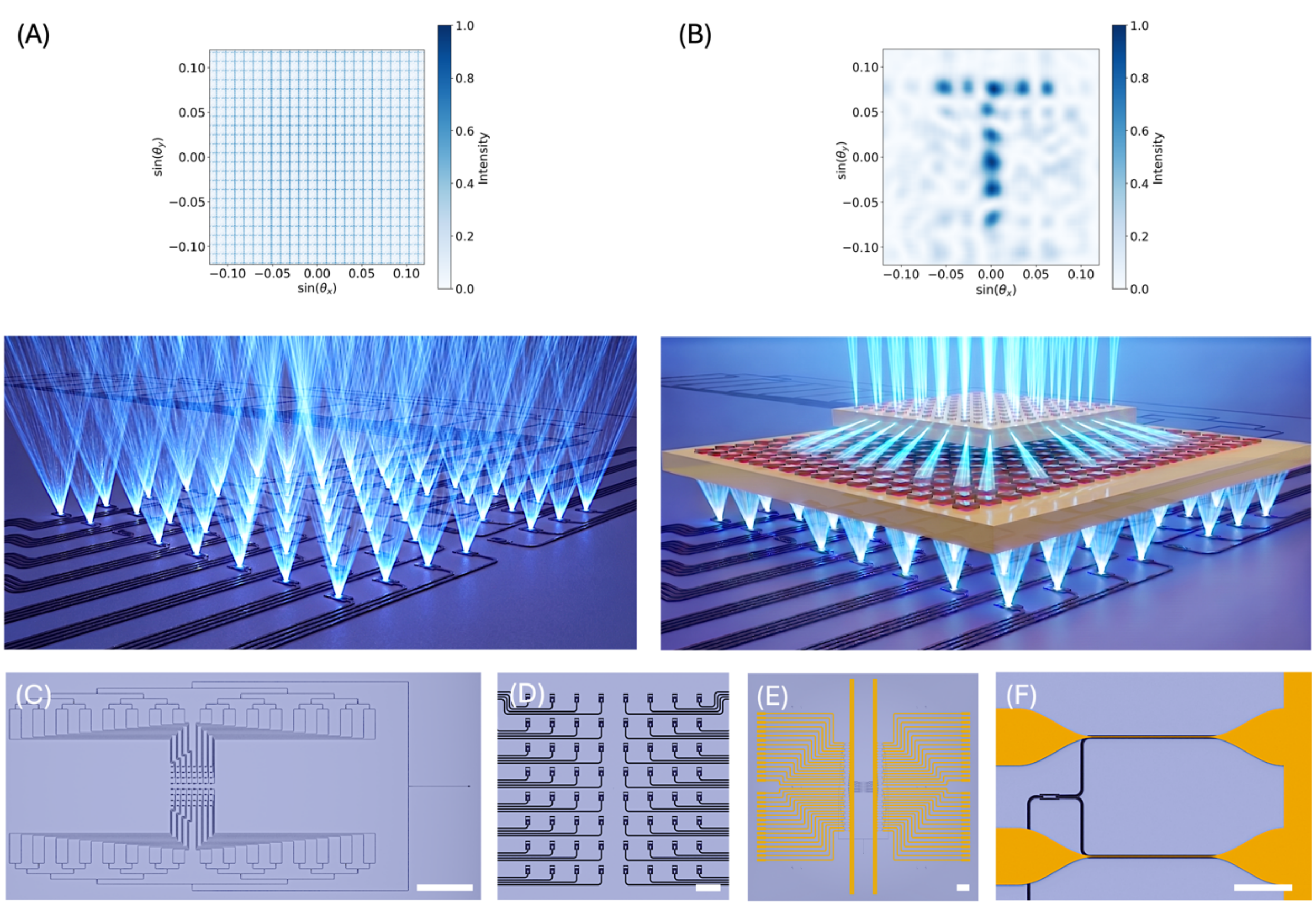

**Figure 1. Device schematic and PIC design.** (A) Simulated far-field pattern showing the letter 'T' alongside higher-order diffraction lobes, generated by an optical phased array with a 150 µm grating pitch without the metasurface doublet. (B) The simulated far-field pattern showing letter 'T' when the grating pitch is effectively reduced to 10 µm using the metasurface doublet (MS1 & MS2). (C) Optical micrograph of the fabricated PIC (Top view), which includes one input grating guiding light to the 64 outputs using a binary tree architecture involving 1×2 MMIs (scale: 1.5

mm). (D) The 8×8 array of output grating couplers as viewed from the top (scale: 150 μm). (E) Top view of the entire PIC, showing the photonic waveguides and electrical wire routing (scale: 1.5 mm). (F) Close-up of the metal heaters used as the thermo-optic phase shifters (scale: 150 μm).

In Fig. 1 we illustrate the schematic of our proposed architecture, comparing the simulated far-field projections both without (Fig. 1A) and with (Fig. 1B) the metasurface doublet. In both configurations, we simulate a far-field holographic projection of the letter 'T'. For the baseline optical phased array (OPA) without the doublet, the far-field is simulated via the Fourier transform of an 8×8 array of equally spaced beams, where the output light from the grating couplers follow a Gaussian profile. Due to the large 150 μm pitch, the far-field of this baseline OPA has a very small FOV of 0.005 (± 0.28°) in numerical aperture (NA) and angle terms as explained in Supplementary Information. We can also see the numerous higher orders of diffraction in Fig. 1A arising from the large pixel pitch making it difficult to decipher the actual symbol. To overcome this, the metasurface doublet effectively achieves a fifteen-fold reduction of the emitter pitch. The first metasurface (MS1) consists of a lenslet array matching the 150 μm pitch of the output grating couplers, steers and focuses the incoming beams onto a second metasurface (MS2) featuring lenslets with a much smaller 10 μm pitch. MS2 then redirects these incoming beams so they travel normal to the PIC plane. We simulate the far-field of this doublet system by first propagating (Angular Spectrum method) the 8×8 Gaussian beams through the phase masks of MS1 and MS2, followed by a Fourier transform of the fields exiting MS2. By effectively demagnifying the beam pitch, the metasurface doublet increases the FOV of the PIC SLM by over 15-fold, from an NA of 0.005 (± 0.28°) to 0.0775 (± 4.44°).

In Fig. 1(C) we show the OPA on a 220 nm silicon-on-insulator (SOI) platform operating at 1540 nm. The PIC employs a binary tree architecture, utilizing MMI couplers to sequentially split the light from a single input grating coupler into 64 distinct output grating couplers. Both the input and output grating couplers are designed to couple light at an 8° off-axis angle. A salient feature of our spatially decoupled architecture is the flexibility to utilize these oblique-emission, large sized grating couplers: the k-vector of the emitted light is readily compensated for by the overlying metasurface. Consequently, the final optical output of the device is perfectly normal to the PIC plane, eliminating the need to design perfectly vertical-emitting grating couplers which often suffer from back reflections. Moreover, the present state of the art highly efficient grating couplers are tens of microns bigger than the wavelength of operation[46], and most previous OPAs had to redesign grating couplers to minimize their footprint[36] [35]. Our architecture can readily employ the large efficient grating, providing high backward compatibility. We show in Fig. 1(D) the top view of the 8×8 output grating coupler array, demonstrating that all elements are uniformly oriented to emit in the same direction. Fig. 1(E) displays the complete layout of the PIC, including the electrical routing required to control the 64 thermo-optic phase shifters. Finally, we present in Fig. 1(F) a magnified view of an MMI coupler, and the platinum (Pt) thermal heaters positioned above the waveguides. The simulation results for the MMI along with its dimensions are given in Supplementary Information. We provide the simulation results for the thermo-optic modulation in Supplementary Information. We use thermo-optic modulation in this work (discussed in Supplementary Information) to demonstrate that despite issues like thermal crosstalk, which require large area between heater filaments to minimize its effect, a full-fledged SLM can be made thanks to our decoupled architecture. Moreover, thermo-optic effect can reach 100 kHz modulation rate[39] which is fast enough to enable several applications beyond current state-of-the-art SLMs.

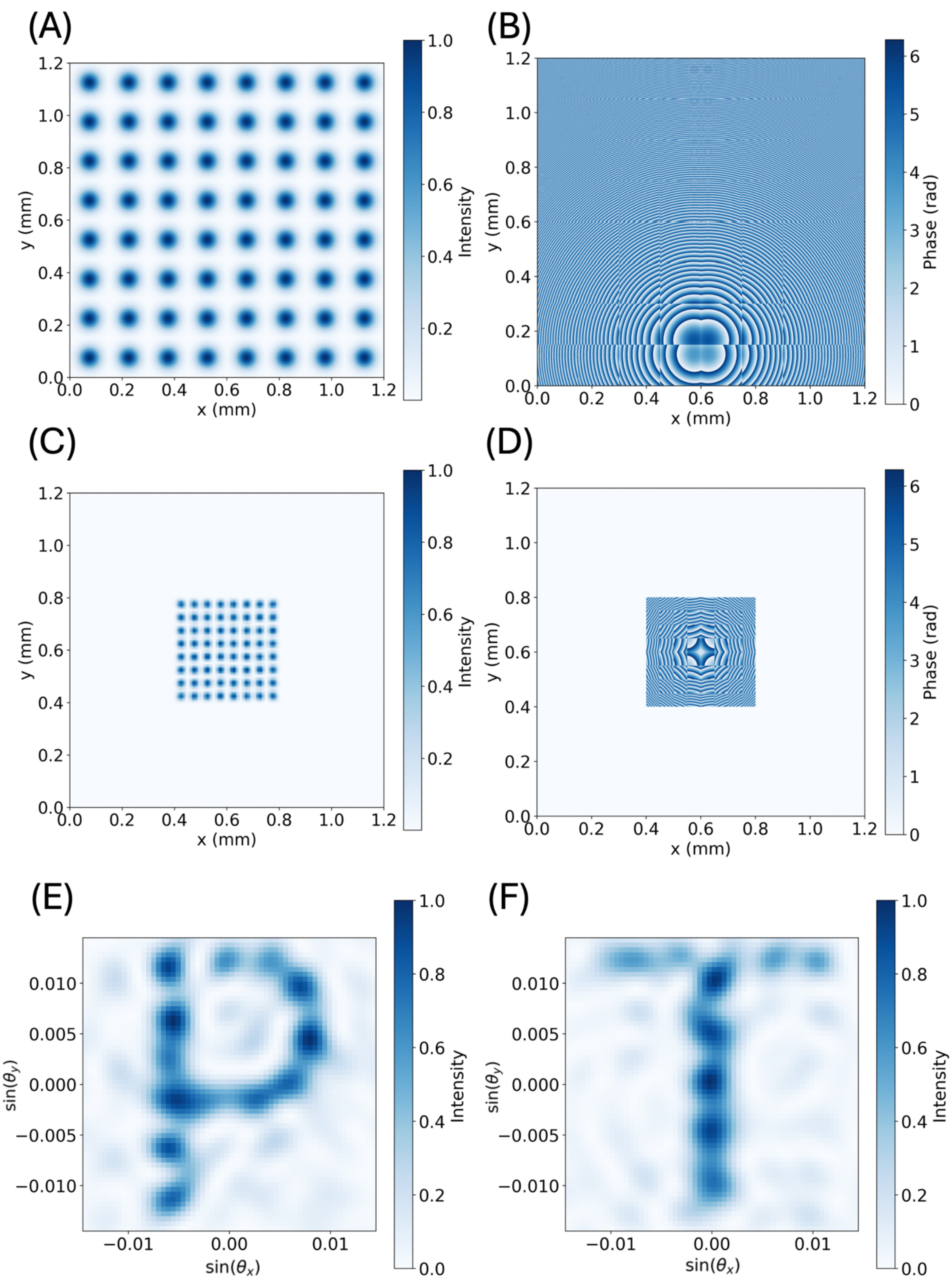

**Figure 2. Metasurface doublet design and simulations** (A) The simulated 8×8 Gaussian beams mimicking the output from the OPA. (B) The simulated phase mask of MS1. (C) The output from MS1, which effectively reduces the beam pitch from 150 μm to 50 μm. (D) The simulated phase mask of the second metasurface (MS2). (E, F) The simulated far-field holograms of letters 'P' and 'T', demonstrating the device's ability to handle both sharp and curved edges with just 64 pixels.

We present the design of the metasurface doublet in Fig. 2. To simulate the metasurface we begin with 8×8 Gaussian beams with a beam waist of 80 μm as presented in Fig. 2 (A). These beams are propagated using the Angular Spectrum method through MS1 with a steering and focusing phase mask plotted in Fig. 2 (B). MS1 contains 64 lenslets arranged in an 8×8 grid with a pixel pitch same as that of the PIC emitters (150 μm). Each lenslet routes the light from their respective emitter to their counterpart in MS2, which also consists of 64 lenslets but with a much smaller pitch (50 μm). We account for the grating coupler's emission angle (8°) to cancel it in the MS1 phase. We note that the absence of rotational asymmetry in MS1 arises due to the oblique emission from the gratings. Detailed description of the MS1 phase mask is given in Supplementary Information. The final location of the focused beams is governed by the pitch of MS2. The MS1 phase profile essentially steers and focuses the incoming beams onto MS2. After passing through MS1, the pitch of the emitted beams is reduced at the plane of MS2, as illustrated in Fig. 2 (C). A second metasurface, MS2, is then used to reverse the steering of the incoming beams so they do not cross over each other in the far field. The phase profile of MS2 is plotted in Fig. 2 (D) and further described in Supplementary Information. Upon exiting MS2, the beams undergo natural Gaussian expansion and interfere in free space. The resulting far-field intensity pattern can be dynamically modulated to project holograms by tuning the phase of the starting Gaussian beams, as depicted in Fig. 2 (E, F). This far-field projection is calculated via Fourier transforming the complex fields immediately after MS2. To demonstrate a proof-of-concept decoupled SLM architecture, we built a system that compresses the pixel pitch 3-fold, from 150 μm to 50 μm, a limit currently dictated by the fabrication constraints of the MS2.

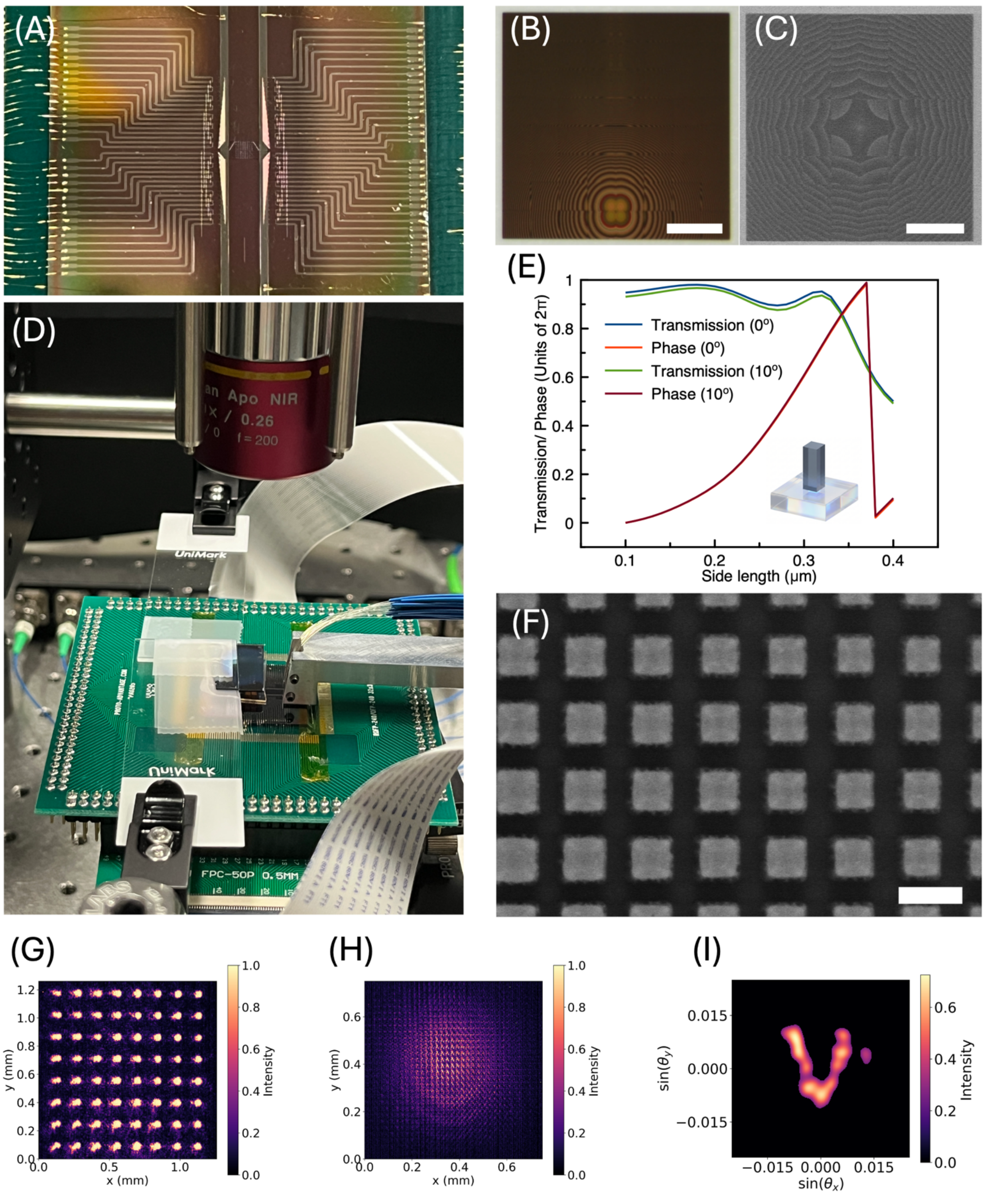

**Figure 3. Fabrication and Experimental Setup** (A) Optical image of the fabricated and wire-bonded PIC chip with metal heaters. (B) Optical micrograph of MS1 (scale: 300 μm). (C) Scanning Electron Micrograph (SEM) of MS2 (scale: 100 μm). (D) The measurement setup with the PIC mounted on a PCB and the metasurface doublet. (E) RCWA simulations for a 1 μm thick Silicon

on Sapphire meta-atom with 500 nm periodicity at 0° and 10° angle of incidence. The side-length of the square meta-atom is changed to implement a varying phase profile in a metasurface. (F) SEM image showing the meta-atoms in the metasurfaces (scale: 500 nm). (G) The infrared image of the initial 8×8 beams. (H) The beams after MS1. (I) The beams after MS2 in the far field interfere to form the hologram of the letter 'V' on application of voltage.

In Fig. 3 we present photographs, optical micrographs, and scanning electron microscope (SEM) images of the fabricated PIC and two metasurfaces. Specifically, Fig. 3(A) displays the PIC wire-bonded to a printed circuit board (PCB) to enable independent electrical control over all 64 heaters (further details are provided in the Methods section). To minimize on-chip electrical routing, the 32 heaters on each side are connected to a common ground. The PCB output transitions from a pin-grid array to a flexible printed circuit (FPC) via an adapter. The opposite end of this FPC cable connects to a commercial voltage controller, which is programmed via Python to supply 64 independent voltages to the heaters. 1540 nm laser light is coupled into the PIC using an 8° fiber. We also provide the zoomed-in micrographs of the PIC and the grating couplers in Supplementary Information.

In Fig. 3(B) we show an optical micrograph of MS1 after fabrication, while Fig. 3(C) presents an SEM image of MS2. Fig. 3(D) displays a photograph of the fully aligned PIC and metasurface assembly, prepared for optical characterization. During assembly, the first metasurface (MS1) is positioned directly above the PIC with a minimal air gap, sufficient to prevent mechanical damage to the grating couplers and meta-atoms. To prevent stray background from light leakage, a metal aperture is fabricated onto the second metasurface (MS2). Fabrication details of the PIC and metasurfaces are reported in the Methods section. MS2 is positioned $\sim 2.5\ mm$ above MS1 where all the beams combine, indicating a slight discrepancy in the fabricated MS1 as compared to the simulated phase mask with $f = 2.2\ mm$. For optical characterization, a near-infrared (NIR) 10× 0.26 NA objective, alongside tube lenses and an infrared (IR) camera, are mounted on a translation stage using an optical cage system. This translation stage enables precise manipulation of the focal plane along the z-axis, facilitating the alignment of all three components (PIC, MS1, and MS2). Further details and schematic of the measurement setup is given in Supplementary Information.

Since MS2 must work at oblique angle of incidence, with a maximum angle of deviation from normal incidence being 10°. Therefore, we simulated 1 μm thick Silicon on Sapphire meta-atom with 500 nm periodicity at both 0° and 10° angle of incidence (Fig. 3E). Thanks to Silicon's high refractive index, our meta-atoms show excellent resilience against a varying angle of incidence. SEM images of the fabricated square meta-atoms are presented in Fig. 3(F). Fig. 3(G) displays an IR camera image of the 1540 nm light emitted from 8×8 grating couplers, captured directly at the PIC plane. Fig. 3(H) illustrates the optical profile captured $\sim 2.5\ mm$ above MS1 at the focal plane where the beams converge. Because the PIC-emitted beams possess a narrower beam waist as compared to the simulation in Fig. 2(A), they diverge rapidly, resulting in a complex interference pattern at the focal point of MS1. Fig. 3(I) presents the far-field intensity profile of the SLM after

MS1 and MS2. By actively tuning the phase of the 64 output couplers, a hologram of the letter 'V' is generated in the far field, recorded with the camera focused on the objective's back focal plane. To optimize this far-field phase profile, we employ an Adaptive Stochastic Parallel Gradient Descent (ASPGD) algorithm[47], which utilizes an adaptive learning rate and real-time feedback. We also compare the far-fields of the PIC with and without the metasurface doublet in Supplementary Information.

The iterative approach implemented via ASPGD algorithm dynamically corrects the inherent random phase variations across the output couplers caused by optical path length differences in the routing waveguides. Furthermore, the algorithm actively mitigates performance degradation stemming from thermal crosstalk, leakage currents, heater resistance variations, and other fabrication-induced imperfections. More details on the phase optimization algorithm are given in the Supplementary Information.

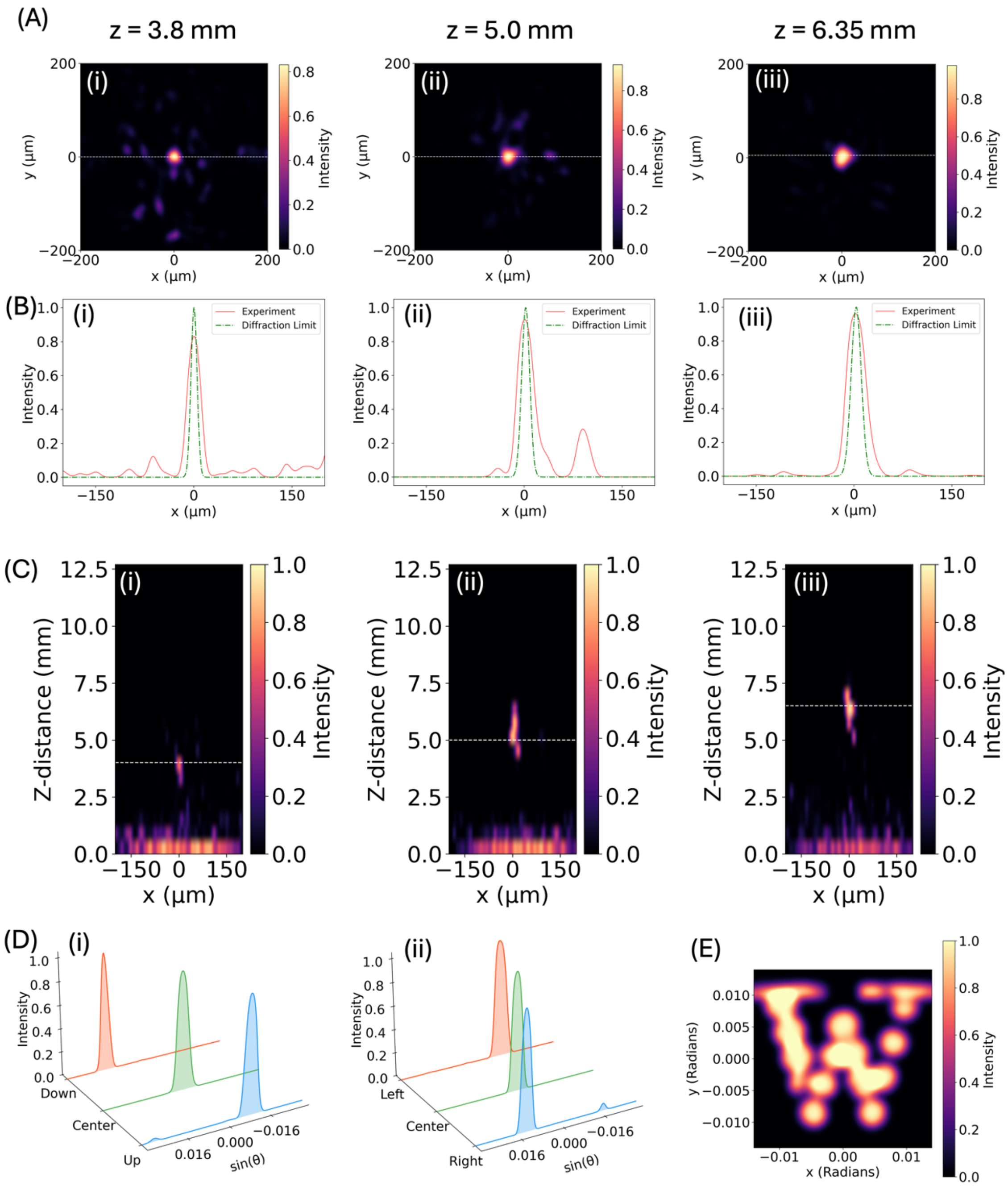

**Figure 4. Tunable varifocal lensing and beam steering.** (A, B, C) (i) Captured images at focal distance above MS2. (ii) The 1D cross-section plot of the focal point. (iii) The z cross-section showing the movement of the focal spot above the chip as we programmed the PIC. (D – i, ii) Plots presenting the measured vertical and horizontal control of the beam. (E) "W" formed using multiple dots and shapes steered at different angles.

Next, we demonstrate the ability to achieve a varifocal lens using our device. In Fig. 4 (A – i, ii, iii), we show the infrared images of the focal spots formed at the z = 3.8 mm, 5.0 mm, 6.35 mm positions, respectively. Fig. 4 (B - i, ii, iii) displays the 1D cross-sections of the focal spots. The full-width-at-half-maximum (FWHM) of the spots is 20 µm, 27 µm and 31 µm respectively, extracted using Gaussian fitting. FWHM of the diffraction limited spot size is 11 µm, 14 µm and 18 µm respectively. We show in Fig. 4 (C – i, ii, iii) the intensity along the optical axis confirming varifocal imaging where z = 0 represents the plane of the MS2. Finally, we show two-dimensional beam steering with both vertical and horizontal beam steering with maximum steering range of ~1.5° (Fig. 4 D, 4E). This amounts to a FOV of 0.013 NA (± 0.74°), which closely matches with the theoretical FOV of 0.015 NA (± 0.86°) for a 50 µm pitch emitter array. Next, we use two-dimensional beam steering to form the word "W" 6.35 mm above MS2, using 14 dots and two rectangular bars (raw images and details of post-processing are given in the Supplementary Information).

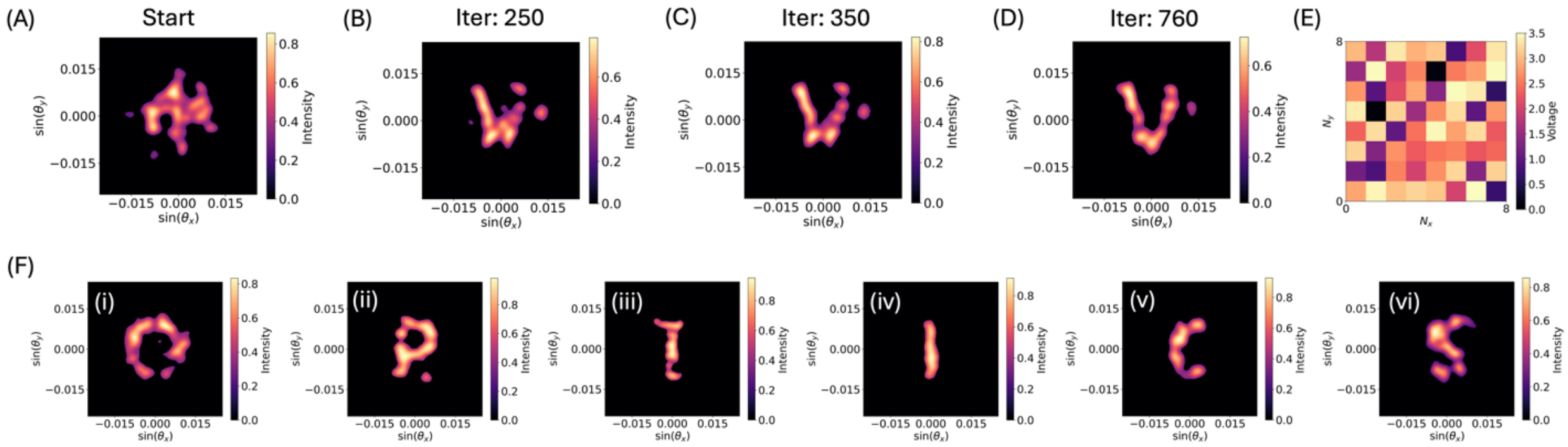

**Figure 5. Two-dimensional holography (**A) The initial far-field intensity profile before starting the optimization loop. (B, C, D) The far-field profiles after 250, 350, and 760 iterations to make the letter 'V'. (E) Heatmap of the applied voltage to the 64 thermo-optic phase shifters at the 760$^{th}$ iteration for forming letter 'V' hologram. (F) Letters 'O', 'P', 'T', 'I', 'C', 'S' created in the far-field.

Finally, we demonstrate two-dimensional holography using our system using the same phase optimization algorithm as before. The schematic of the optical setup used to carry out this measurement is given in Supplementary Information. Fig. 5(A) displays the initial far-field intensity profile resulting from a random voltage applied across the 64 thermal heaters. To illustrate the evolution of the hologram during optimization, the far-field intensity patterns at iterations 250, 350, and 760 are presented in Figures 5(B–D) for the target letter 'V'. Fig. 5(E) presents the applied voltage heatmap (varied between 0 to 3.5V) at the 760$^{th}$ (final) iteration for the 64 thermo-optic phase shifters.

More details of the phase optimization procedure, post-processing, alongside the raw images, are provided in the Supplementary Information. Ultimately, Fig. 5 (F, i–vi) showcases the consecutive generation of multiple holographic letters to spell the word "OPTICS," successfully validating the device's capability for reconfigurable 2D holography.

## Discussion

We have proposed and experimentally demonstrated a new architecture for SLMs that overcomes the fundamental spatial constraints of electrical routing by decoupling spatial and temporal degrees of freedom. By physically separating the electrical routing/ the PIC modulation plane (temporal plane) from the final optical output plane (spatial plane), we eliminated the stringent requirement of confining complex electronics within a sub-wavelength pitch. Utilizing a metasurface doublet integrated directly atop a 64-channel silicon PIC at 1540 nm, we achieved a 3-fold reduction of the pixel pitch (from 150 µm to 50 µm). The pixel pitch reduction further expanded the system's FOV by ~3-fold, increasing the numerical aperture from 0.005 (± 0.28°) to 0.013 (± 0.74°). Furthermore, via a gradient descent algorithm with real-time feedback, we effectively mitigated inherent phase errors arising from thermal crosstalk, and fabrication imperfections. To the best of our knowledge, this work represents the first simultaneous demonstration of varifocal lensing, dynamic beam steering, and two-dimensional tunable phase holography on a PIC-based SLM using just 64 emitters. Using PICs to take advantage of their tight light confinement and agility of metasurfaces to route light in free space we bring the best of both worlds' together to enable a hybrid architecture. This new SLM architecture provides a highly scalable roadmap for realizing large-FOV, high-speed, and low-power reconfigurable optics. Looking forward, the natural progression of this decoupled architecture is to scale the SLM to a significantly larger pixel count while simultaneously compressing the pixel pitch into the deep sub-wavelength regime. The future work can be broken down into three parts:

First, achieving a true sub-wavelength pitch is critical for entirely suppressing higher-order diffraction lobes (Supplementary Information). While the current metasurface doublet successfully serves as a proof-of-concept for demagnifying the pitch from 150 µm to 50 µm, further reduction to the sub-wavelength regime introduces fundamental optical challenges. Specifically, extreme downscaling pushes the MS2 lenslets against the fundamental footprint limit required for a structure to function as a focusing lens rather than a simple aperture. At an operating wavelength of 1550 nm and a focal length of 100 µm, the lens has to be at least 15 × 15 µm² in size for it to have a Fresnel number greater than 1 [48]. Although the focal length could theoretically be shortened to allow for a smaller aperture, this approach quickly encounters fabrication limits as adequately sampling the phase profile requires a minimum of two meta-atoms per Fresnel zone [48].

Second, scaling up the total pixel count inherently enlarges the aperture of MS1. Coupling this expanded MS1 aperture with a compact MS2 forces the peripheral regions of MS1 to steer light at increasingly oblique angles. For instance, expanding the emitter array from 8 × 8 to 16 × 16 while maintaining a constant focal length and pitch almost doubles the required steering angle at the periphery lenslets. Even if MS1 lenslets successfully supports these extreme deflection angles, maintaining a consistent numerical aperture (NA) across the varying focal lengths of MS1 becomes highly problematic. Because the effective focal length of the peripheral lenslets is significantly longer than that of the central lenslets, the resulting focused spot sizes illuminating MS2 will suffer from severe spatial non-uniformity. Furthermore, the light arriving from the enlarged MS1 aperture will strike the peripheral regions of MS2 at extreme angles of incidence, strictly necessitating the use of highly angle-insensitive meta-atoms[49]. To circumvent these flat-optic scaling limits entirely, future architectures could replace the metasurface doublet with volume optics, such as 3D-printed photonic lanterns[50], seamlessly routing light from the integrated photonic circuit to the output plane.

Third, as we push modulation speeds into the GHz regimes by utilizing electro-optic modulation, the overall dynamic power consumption of the SLM will scale linearly with the operating frequency (Supplementary Information). However, PICs, owing to their ability for tight optical confinement, are well positioned to overcome this bottleneck using well established methods such as high-Q micro-ring integrated electro-optic modulators[51]. While we used university cleanroom to fabricate the PIC and metasurfaces, they can be easily scaled using commercial foundries[52].

We note that the phase optimization procedure used in this work is time consuming however this can be easily replaced with one time calibration by implementing on-chip interferometers[53].

**Methods**

PIC fabrication: The PIC was fabricated on a 220 nm Silicon (Si) on Insulator (SOI) 25x25 $mm^2$ chip. The photonic circuit was patterned on the SOI sample using ZEP resist and 100kV E-beam lithography (JEOL 6300FS). After E-beam lithography the pattern was transferred onto Si using Reactive-Ion Etching (RIE) with fluorine gas (Oxford Plasma Lab 100). ZEP was then stripped off using Acetone and IPA sonication. To shield the PIC from metal losses 1 μm of $SiO_2$ was deposited using PECVD at 350°C (SPTS SPM). The metal heaters and metal wiring were patterned on top of the $SiO_2$ layer using NR9G-3000PY negative resist exposed using photolithography (Heidelberg DWL66+). 20nm Titanium (Ti) adhesion layer and 300nm Pt were then deposited using E-beam evaporator (Evatec LLS EVO). Metal lift-off was then performed by gentle sonication followed by overnight Acetone submersion of the sample. Finally, the PIC was wire bonded (WESTBOND 4700E) onto the main PCB (Proto Advantage PA0205). The main PCB is the connected to a FPC adapter (FPC-50P 0.5mm).

Metasurface fabrication: Both the metasurfaces were fabricated on 1 μm Si on Sapphire wafers. The metasurface was patterned using ZEP resist and 100kV E-beam lithography (JEOL 6300FS). Following which a hard mask of Alumina ($Al_2O_3$) was deposited using E-beam evaporation (CHA SEC-600). The hard mask was then lifted off using overnight NMP submersion. Lastly, Si was etched to transfer the pattern from the hard mask to Si using RIE etching with fluorine gas. Metal aperture was fabricated on top of the second metasurface using NR9G-3000PY negative resist exposed using photolithography (Heidelberg DWL66+).

Electrical control: Eight Qontrol voltage drivers (Q8b) each with 8 individually controllable voltage outputs were used to apply reconfigurable voltage to the 64 thermo-optic phase shifters. The Q8b's were attached to the main board (BP12) which was then attached to an intermediatory board (INT12FFC2) to facilitate electrical signal routing to the sample using two 50-pin FCC cables.

Optical setup: A 10x 0.26 NA Mitutoyo NIR objective along with 50 mm and 100 mm lenses were used to image the device onto an infrared camera (Pembroke SenS SWIR). The objective along with the lenses and camera were placed vertically onto a Mitutoyo translational stage. A Santec TSL-570 laser was used to supply the 1540 nm laser input. Schematic of the setup is given in Supplementary Information.

## Acknowledgment

This work is supported by NSF-FuSe and NASA-STTR awards. Part of this work was conducted at the Washington Nanofabrication Facility / Molecular Analysis Facility, a National Nanotechnology Coordinated Infrastructure (NNCI) site at the University of Washington, with partial support from the National Science Foundation via awards NNCI-1542101 and NNCI-2025489.

## Author Contributions

AM and VT conceptualized the project. VT designed the PIC layout, simulated the MMI and grating couplers. AWS designed and wrote the code to simulate the metasurfaces. VT simulated the metasurfaces and performed hologram generation simulations. VT fabricated the metasurfaces, photonic integrated circuit and did the device packaging. VT performed the device measurements. AM and JF supervised the progress of the project. VT wrote the manuscript with inputs from all the authors.